\documentclass[twocolumn,amssymb]{revtex4}
\usepackage{epsfig}
\usepackage{hyperref}

\begin{document}
\title{ The study of birefringent homogenous medium with geometric phase}
\author{Dipti Banerjee}
\thanks{Regular associate of ICTP}
\email{deepbancu@homail.com, dbanerje@ictp.it}
\affiliation{Department of Physics \\
Vidyasagar College for Women \\
39, Sankar Ghosh lane,Kolkata-700006,\\ West Bengal, INDIA}
\affiliation{\bf{The Abdus Salam International Center for Theoretical
Physics},\\ Trieste,ITALY}
\date{30.06.10}

\begin{abstract}
The property of linear and circular birefringence
 at each point of the optical medium has been evaluated here from differential matrix $N$ using the Jones calculus.This matrix lies on the OAM sphere for $l=1$ orbital angular momentum.The geometric phase is developed by twisting the medium uniformly about the direction of propagation of the light ray. The circular birefringence of the medium,is visualized through the solid angle and the angular twist per unit thickness of the medium, $k$,that is equivalent to the topological charge of the optical element.
\end{abstract}

\maketitle
keywords: Birefringent,geometric phase, topological charge.\\
Pacs code: 42.25.Bs

The cyclic transformations of polarized light develops both dynamical and geometric phase (GP).This GP has attracted attention in the last 25 years for its robustness against the dynamical change of the environment,insensitive to vibrations and other mechanical effects. Four manifestations of GP have been reported in optics so far.
i) Pancharatnam phase \cite{paper1} is $\Omega/2$, the first identified GP originated from the cycle change of polarizations of a plane polarized light having fixed direction over a closed path subtending the solid angle $\Omega$ on the Poincare sphere.Berry found the quantal counterpart \cite{paper2} of Pancharatnam's phase in case of cyclic adiabatic evolution.He also studied the phase two-form (GP) \cite{paper3} in connection with the dielectric tensor and
birefringence of the medium.
ii) The second kind of phase was experimentally performed by Chaio and co-workers \cite{paper4} when the light with fixed polarizations slowly rotate around a closed circuit with the varied directions. The developed GP was the spin-redirection or coiled light phase.
iii) The third one was developed by the squeezed state of light through the cyclic changes of Lorentz transformation \cite{paper5}.
iv) The fourth GP was studied by van Enk \cite{paper6}, in case of cyclic change of transverse mode pattern of Gaussian light beam without affecting the direction of propagation or the polarization of light.

To study different types of Geometric phases,the polarization or direction of propagation of the polarized light has to be varied over a closed path.This could be possible by passage of polarized light through optical elements such as linear and circular polarizer, retarder, rotator,etc arranged in proper sequence.The works of Jones \cite{paper7},\cite{paper8},attracted our attention for the formulation of optical transformation in terms of $(2\times2)$ matrix.
Azzam studied \cite{paper9} the passage of polarized light through anisotropic medium both with and without depolarization,using differential matrix of Stokes and latter Jones.
The idea of Jones that the plane polarized light after passing through a rotator suffers a rotation of plane of polarization has been re-established by us \cite{paper10}
from the view point of geometric phase.Explicitly we considered \cite{paper11} that the photons in a polarized beam fixes the helicity whose direction changes with the change of polarization. With the spinorial representation polarized photon by spherical harmonics,
 the dielectric property of the anisotropic medium \cite{paper12} has been studied further from the view point of geometric phase.

The physical mechanism of these different kind of GPs in optics originate from spin or orbital angular momentum of polarized photon.Now a days the angular momentum of photon plays an interesting role both in the theoretical and experimental studies.Indeed,Enk studied GP in mode space and pointed out \cite{paper6} that Pacharatnam phase is associated with spin angular momentum transfer of light and optical medium resulting the exchange of momentum inevitable.The direction of spin changes in the second kind of phase,and GP (for LG beams) possesses orbital angular momentum due to transverse spatial dependence.The first observation of the angular momentum of light was performed by Beth \cite{paper13} through an experiment where a beam of right circularly polarized light was passed through a birefringent medium (quarter-wave plate) and transformed to left circularly polarized light.Galvez et.al. gave an experimental measurement of GP \cite{paper14} in mode space acquired by the orbital angular momentum.In the interaction between radiation and matter,Padgget \cite{paper15} showed his interest on the orbital angular momentum of polarized photon in Poincare sphere representation.The equivalence between the phase shift introduced by birefringent wave plates and mode converter allows one to treat the mode structure formulation analogous to the Jones matrix. There is an interesting analogy \cite{paper16} between rotations in one formulation and mode converters in another formulation. Tiwari \cite{paper17} reviewed in detail the connection of different kinds of GPs in optics with OAM and SAM.

Berry \cite{paper3} studied the phase two-form (GP) of an optical medium possessing both birefringence and gyrotropy.Motivated by his work we here will focus on the non-absorbing but pure birefringent anisotropic optical medium represented by differential matrix $N$.  Following Jones \cite{paper8} work, the appearance of GP could be studied here in two ways by.\\
   i) uniformly twisted crystal.\\
   ii) arbitrarily twisted crystal.\\

 For a particular polarization,light has fixed helicity whose direction changes with the change of polarization.Thus the variation of polarization over a closed path, GP could be studied \cite{paper10},\cite{paper11} and \cite{paper12} in connection with the helicity of polarized photon.The variation of the direction of the polarized light passing through twisted optical medium would cause an exchange of the optical power between the two states of the light along two directions for which a natural twist is visible.In this communication, we have in mind to give an angular momentum interpretation of GP appeared by twisting a pure birefringent optical medium.

   A Jones matrix representation of birefringent medium will be given in the section-1, study of geometric phase in section-2 and lastly the angular momentum interpretation of the GP appeared in present and previous works in the last section-3.

\section{The Jones matrix representation of birefringent medium}
A light beam is said to be polarized whenever it is transmitted through a certain crystalline medium that allows electrical anisotropy \cite{paper7}.This change of polarization state can be written as
\begin{equation}
\varepsilon=M \varepsilon_o
\end{equation}
where $\varepsilon$ and $\varepsilon_o$ are the respective final and initial polarization state and M is polarization matrix respectively.
If the polarization of light remains unaltered after passing through any optical element then the state can be identified as the eigenvector of the optical component and in the language of matrix,Jones had shown the condition
\begin{equation}
M_i \varepsilon_i=d_i\varepsilon_i
\end{equation}
 where $d_i$ is the eigenvalue corresponding to
  the eigenvectors $\varepsilon_i$ of a particular polarization matrix
 $M_i=\pmatrix{m_1&m_4\cr m_3&m_2}.$\\

 The optical properties such as birefringence and dichroism of a homogeneous medium varies with distance.The passage of light through an optical element such as birefringent, absorbing or dichroic plate would be to change both $d_x$ and $d_y$ so that the effect may be represented by $|\psi_f>=M|\psi_i>$.
For a non-absorbing plate, there is no change in the intensity and the polarization matrix $M$ is therefore unitary $detM=1$ which makes $|\psi_f|=|\psi_i|$.
  and can be studied by the differential matrix $N$ which refers to the optical element for a given infinitesimal path length within the element.For particular wavelength and direction,the evolution of a light vector $\varepsilon$ becomes
\begin{equation}
\frac{d\varepsilon}{dz}=\frac{dM}{dz}\varepsilon_0
=\frac{dM}{dz}M^{-1}\varepsilon=N\varepsilon
\end{equation}
where it is evident that $N$ is the operator that determines $dM/dz$ from $M$ as follows
\begin{equation}
N=\frac{dM}{dz}M^{-1}=\pmatrix{n_1 & n_2 \cr n_3 & n_4}
\end{equation}
When $N$ is independent of z, then on integration the dependence of polarization matrix M on z is seen from
\begin{equation}
M=M_0\exp(\int{Ndz})
\end{equation}
Jones had shown further that the eigenvectors of M are equal to that of N when it is independent of z \cite{paper8}.

Any homogeneous crystal without optical activity could be considered for normal incidence as a laminated crystal.According to the lamellar representation suggested by Jones \cite{paper7},a thin slab of a given medium is equivalent to a pile of retardation plates and partial polarizers.Eight constants are required to specify the real and imaginary parts of the four matrix elements of $2\times2$ N matrix, each possessing one and only one of the eight fundamental properties.The eight optical properties are paired \cite{paper9} and reduce to four.\\
i) Isotropic refraction and absorption\\
ii) Linear birefringence and linear dichroism along the xy coordinate axis.\\
iii) Linear birefringence and linear dichroism along the bisector of xy coordinate axes.\\
iv) Circular birefringence and circular dichroism.\\
The optical medium that has circular birefringence and linear birefringence will be our point of interest, and could have the following matrix form
\begin{eqnarray}
\theta_{cb}= \eta \pmatrix{0 & -1 \cr 1 & 0} \\
\theta_{lb}= \rho \pmatrix{0 & -i \cr i & 0}
\end{eqnarray}
These $\theta_{cb}$ and $\theta_{lb}$ matrices form the required differential matrix.
\begin{eqnarray}
N=\theta_{cb}+\theta_{lb}
=\pmatrix{0 & -\eta+i\rho \cr \eta+i\rho& 0}=\pmatrix{0 & n_2 \cr n_3 & 0}
\end{eqnarray}
where $\eta$ is the circular birefringence that measures the rotation of the plane polarized light per unit thickness and
$\rho$ is the part of linear birefringence that measures the difference between the two principal constants along the coordinate axes.

A crystal will be considered homogeneous \cite{paper18}, if any one of the optical property will be visible instead of eight (as if all the eight lamina are sandwiched).The evolution of the ray vector $\varepsilon={\varepsilon_1 \choose \varepsilon_2}$ when passes through such medium $N$,as in eq.(3) could be re-written into the components as
\begin{eqnarray}
\frac{d\varepsilon_1}{dz}=n_{1}\varepsilon_1 + n_{2}\varepsilon_2\\
\frac{d\varepsilon_2}{dz}=n_{3}\varepsilon_1 + n_{4}\varepsilon_2
\end{eqnarray}
For pure birefringent medium represented by eq.(8), one may use the evolution of ray vector
\begin{equation}
\frac{d\varepsilon_1}{dz}=n_2 \varepsilon_2,\\
\frac{d\varepsilon_2}{dz}=n_3 \varepsilon_1
\end{equation}
This shows that as the light enters into the birefringent plate,the spatial variation of component of electric vector in one direction gives the effect in the other perpendicular direction. It means that there is an exchange of optical power between the two component states of the polarized light indicating the rotation of the ray vector after entering the medium.

Geometrically this state $\varepsilon$ is a point P on the surface of the Poincare sphere that defines a position vector $\vec{p}$ in three dimensional space. The evolution of the vector $\vec{p}$ is equivalent to the cyclic change of the state vector during the passage of infinitesimal distance dz of the optical medium. Huard pointed out in his book \cite{paper18} that the spatial change of vector as it passes through the crystal becomes
\begin{equation}
\frac{d\vec{p}}{dz}=\vec{\Omega} \times \vec{p}
\end{equation}
 This shows a natural twist by the instantaneous rotation vector $\Omega$ along the axis oz about which the $\vec{p}$ makes an elementary angle $d\alpha=\Omega dz$, for thickness $dz$. The magnitude and direction of the rotation vector depends on the inherent property of the medium, in other words on the element of the N matrix.

(i) {\bf The uniformly twisted crystal}\\

When an originally homogeneous crystal twisted uniformly about an axis parallel to the direction of transmission, the $N$ (z dependence) matrices are transformed upon rotation where $k$ is the angular twist of the optical medium per unit thickness \cite{paper8}.
It could be noted here that this $k$ has similar definition of $\Omega$ as
 in eq.(12), having the basic difference in their space of appearance.
 In fact,$k$ is associated with the external rotation of the optical element
 where as $\Omega$ is the instantaneous rotation of the polarized light incident on the optical element by an angle.

 The differential matrix $N$ which corresponds to the twisted crystal is related with its untwisted position matrix $N_0$ by the following relation
\begin{equation}
N=S(kz)N_0 S(-kz)
\end{equation}
where $S$ is the rotation matrix having developed after uniform rotation about an angle $\theta=kz$.
Jones realized the simultaneous rotation of the twisted state $\varepsilon'$ in the opposite direction of N matrix,
\begin{equation}
\varepsilon'=S(-kz)\varepsilon
\end{equation}
so that it satisfies the following equation
\begin{equation}
\frac{d\varepsilon'}{dz}=N'\varepsilon'
\end{equation}
with the dependence on the twisted matrix $N'$ that has independence on $z$.
It has been shown \cite{paper7} after few steps that this twisted matrix $N'$ can be expressed
\begin{equation}
 N^\prime=N_0 - kS(\pi/2)
\end{equation}
in terms of $N_0=$ the matrix for untwisted crystal and $S(\pi/2)$ denotes the rotation matrix for normal incidence of light.
The solution of the above eq.(15) may be written $\varepsilon'=\exp(N'z)\varepsilon'_0$
where $\varepsilon'_0$ is the value of the vector $\varepsilon'$ at $z=0$.

(ii) {\bf The arbitrarily twisted crystal}\\

The transformation illustrated above may also be applied with an arbitrarily twisted crystal.Jones showed \cite{paper7} that the $N$ matrix of the twisted crystal is
\begin{equation}
N=S(\omega(z))N_0 S(\omega(z))
\end{equation}
where $\omega(z)$ specifies the angle of twist, that is the arbitrary function of z.
From the similar transformation of the uniformly twisted crystal, one finds
\begin{equation}
N'=N_0-(\frac{d\omega(z)}{dz})S(\pi/2)
\end{equation}
It has been pointed out by Jones that the derivative $\frac{d\omega(z)}{dz}$ is a constant for a uniformly twisted crystal.
Due to the inherent property of the birefringent optical medium (N)
a natural twist is realized by the incident polarized light. Further, external twist of the medium might develop an additional phase in fixed OAM sphere.
We realize this phase in the next section as OAM holonomy which will visualize the circular dichroism of the medium.

\section{The geometric phase by twisting homogeneous medium}

The propagation of light through an optical system was studied by Jones \cite{paper7}
in a $2x2$ matrix method.The method was based on the idea that in anisotropic media,the displacement vector D can be represented by a column vector.On similar manner Berry \cite{paper2} pointed out that a monochromatic light traveling in the z direction,the polarization state of the electric displacement vector lying in the xy plane can be written by a two component spinor
$$|\psi>={\psi_+ \choose \psi_-}$$
 where  $\psi_{\pm}=(d_x \pm i d_y)$ and the intensity is $I=|d_x|^2+|d_y|^2$ while the complex ratio $d_x/d_y$ defines its polarization state.

 The polarization matrix $M$ satisfying $M|\psi>=1/2 |\psi>$ can be determined from the eigenvector $|\psi>$ with eigenvalue $+1/2$ using the relation $(|\psi><\psi|-1/2)$.If we consider the eigenvector by \cite{paper10}
\begin{equation}
|\psi>={\cos\theta/2 e^{i\phi} \choose \sin\theta/2}
\end{equation}
the polarization matrix in terms of polar angles $\theta$ and $\phi$ represent different point on the surface of the Poincare sphere,
\begin{equation}
\begin{array}{lcl}
M=1/2 \pmatrix{\cos\theta & \sin\theta e^{i\phi} \cr \sin\theta e^{-i\phi} & -\cos\theta \cr}
\end{array}
\end{equation}
having eigenvalues $\pm 1/2$.This eigenvalues reflect the helicity of polarized photon that has been used \cite{paper11}\cite{paper12} to evaluate the respective GP.
Here the parameters $\theta$ and $\phi$ are the coordinates of polarized light that represent a point on the Poincare sphere.

In search of the birefringent plate at a particular position $z$ the differential matrix $N$ of optical medium has been calculated from the polarization matrix $M$ as in eq.(20). Since the eigenvalues of helicities for polarized photons is
$\pm1$, we omitted the factor $1/2$ from the polarization matrix $M$ \cite{paper11} in the following equations.
\begin{equation}
N=(\frac{dM}{d\theta})(\frac{d\theta}{dz})M^{-1}
\end{equation}
 Considering the angle of twist being proportional directly to the thickness z of the optical medium, $\theta=kz$ where $k$ represent the angular twist per unit thickness,
 the corresponding differential matrix N \cite{paper11} becomes
\begin{equation}
N=k \pmatrix{0 & -e^{i\phi} \cr e^{-i\phi} & 0 \cr}
\end{equation}
having complex eigenvalues $\pm ik$ with eigenvectors
\begin{equation}
\left(
\begin{array}{c}
 \pm i e^{i \phi } \\
 1
\end{array}
\right)
\end{equation}

The nature of the optical medium could be identified
comparing the above N matrix in eq.(22) with eq.(8).It is seen that our N matrix is homogenous and possesses both circular and linear birefringence by $k\cos\phi$ and $(-k\sin\phi)$ respectively.It may be noted that in case one observes the eigenvalues of N opposite and imaginary,the optical medium possesses the property of purely circular birefringence \cite{paper7}.

Initially we consider the angular twist per unit thickness zero, $k=0$ (for $\theta=0$), then for a uniformly twisted optical medium, using eq.(16),
the twisted matrix becomes
\begin{equation}
N'=\pmatrix{0 & k \cr -k & 0 \cr}
\end{equation}
where the untwisted matrix $N_0=0$ and the rotation matrix
$S(\pi/2)=\pmatrix{0 & -1 \cr 1 & 0 \cr}$. The twisted ray will be obtained after the opposite rotation of the twisted matrix from eq.(14)
\begin{equation}
\varepsilon'=\pmatrix{\cos\theta & \sin\theta \cr -\sin\theta & \cos\theta \cr}
\left(
\begin{array}{c}
 i e^{i \phi }\\
 1
\end{array}
\right)
\end{equation}
in other words
\begin{equation}
\varepsilon'=
\left(
\begin{array}{c}
 ie^{i\phi}\cos\theta+\sin\theta \\
 -ie^{i\phi}\sin\theta+\cos\theta
\end{array}
\right)
\end{equation}.

 The initial eigenvector $\varepsilon$ reappears if opposite rotation is given to $\varepsilon'$. Now considering the matrix in eq.(24) as initial matrix $N_0$,
 by using $k=1$ which results the twisted matrix $N'$
 \begin{equation}
 N'= \pmatrix {0 & -e^{i\phi}+k \cr  e^{-i\phi}-k & 0 \cr}
 \end{equation}
It can be realized looking at eqs.(22),(24) and (27) that $k$, the angular twist per unit thickness in the birefringent optical medium plays the similar role as $\Omega$ for twisting a ray vector as in eq.(12).

Light having fixed polarization and helicity, if suffers the slow variation of path in real space it can be mapped on to the surface of unit sphere in the wave vector space.
The geometric phase is found to appear as the initial state $|A>$ unite with final $|A'>$.
\begin{equation}
<A|A'>=\pm \exp(i\Omega(C)/2)
\end{equation}
where $\Omega$ is the solid angle swept out by $e_k$ on its unit sphere.

Our present work is based on the consideration of polarized light passing normally
through a medium N having linear and circular birefringence. Due to the inherent property of the medium, the incident polarized light suffers a natural twist about the axis parallel to the direction of its propagation. We assume $\Upsilon$ and $\Upsilon'$ are the respective phases developed as the respective initial states $|A>=\varepsilon$ or $\varepsilon'$, are passed through the respective differential matrices $N$ or $N'$
\begin{eqnarray}
\Upsilon=\varepsilon^* \frac{d\varepsilon}{d\theta}\frac{d\theta}{dz}={\varepsilon^*}N\varepsilon\\
\Upsilon'={\varepsilon'}^* \frac{d\varepsilon'}{d\theta}\frac{d\theta}{dz}={\varepsilon'}^* N'{\varepsilon'}
\end{eqnarray}
with the consideration of $d\theta=d\theta'$.

$\Upsilon$ can be obtained,using eq.(22) and (23) and (11) in (29) in the following equation
\begin{eqnarray}
\Upsilon={\varepsilon^*}N\varepsilon
&=&{\varepsilon_1}^*n_2\varepsilon_2+{\varepsilon_2}^*n_3\varepsilon_1\nonumber\\
&=&(-ie^{-i\phi})(-ke^{i\phi})+(ke^{-i\phi})(ie^{i\phi})\nonumber\\
&=&2ik
\end{eqnarray}
Hence the phase of untwisted medium becomes $2ik$. By varying the value of $k=0,1,2..etc$, different $\Upsilon$ can be obtained.
 In a similar way for twisted medium, we use the twisted matrix $N'$ eq.(24) for $k=0$ and the initial state by eq.(26) to evaluate $\Upsilon'$.
\begin{eqnarray}
\Upsilon'&=&{\varepsilon'}^*N'{\varepsilon'}
={{\varepsilon'}_1}^*{n'}_2{\varepsilon'}_2+{\varepsilon_2}^* {n'}_3 {\varepsilon'}_1\nonumber\\
&=&(-i e^{-i\phi}\cos\theta+\sin\theta)k(-ie^{i\phi}\sin\theta+\cos\theta) \nonumber\\
&+&(ie^{-i\phi}\sin\theta+\cos\theta)(-k)(ie^{i\phi}\cos\theta+\sin\theta)\nonumber\\
&=& -2ik\cos\phi
\end{eqnarray}
Any ray passing through $N$ or $N'$ will suffer a twist due to the internal dynamics of the birefringent medium. The polarized light passing through the twisted medium $N'$ will acquire the phase $\Upsilon'$,that has two parts acquired one from the dynamics and another parametric change of the medium.Thus the phase $\Upsilon'$ will contain both the dynamical and geometric phase of a uniformly twisted birefringent medium.
To grasp the geometric phase of the incident polarized state due to external twist of a birefringent medium in an initial adjustment $k=0$, we
take the difference between the two phases $\Upsilon'-\Upsilon$ that will eliminate the dynamical phase, the phase due to natural twist, resulting the outcome of geometric phase for normal incidence of the polarized light on the medium.

As a result the required geometric phase becomes
\begin{equation}
\Upsilon'-\Upsilon=-2ik\cos\phi
\end{equation}
where it is seen $\Upsilon=0$ for $k=0$.
On similar manner the required geometric phase of the other conjugate eigenvector will be
\begin{equation}
\Upsilon'-\Upsilon=2ik\cos\phi
\end{equation}
Thus it is seen the geometric phase visualize the circular birefringence of the medium by $k\cos\phi$.

If the orientation of the matrix N is changed considering $k=1$, then using the initial matrix\\
$N_0=\pmatrix{0 & -e^{i\phi} \cr e^{-i\phi} & 0 \cr}$
one can find the phase $\Upsilon=\pm 2i$ for the respective eigenvectors in eq.(23).

To calculate the corresponding twisted phase $\Upsilon'$, the twisted matrix $N'$ in eq.(27) is used that intuitively will act on the twisted light ray $\varepsilon'$ in opposite direction for making $k=1$.As result the original eigenvector $\varepsilon=\varepsilon"=S(\theta)\varepsilon'$ is visible.
In the language of mathematics the twisted GP becomes
\begin{eqnarray}
\Upsilon'&=&{{\varepsilon}_1}^*{n'}_2{\varepsilon}_2+{\varepsilon_2}^* {n'}_3 {\varepsilon}_1\nonumber\\
&=&(-i e^{-i\phi}, 1)\pmatrix {0 & -e^{i\phi}+k \cr  e^{-i\phi}-k & 0 \cr}\\
\left(\begin{array}{c}
 i e^{i \phi }\\
 1
\end{array}
\right)
&=& 2i[1-k\cos\phi]
\end{eqnarray}
This helps us to recover again the previous form of the geometric phase
\begin{equation}
\Upsilon'-\Upsilon =2i[1-k\cos\phi]-2i= -2i(k\cos\phi)
\end{equation}
 Hence again the GP in terms $k\cos\phi$ is visible where the non-zero $k$ is associated with the particular twist of the crystal. We choose the light incident at a particular angle $\theta$ on the optical medium. Now for the first choice $k=0$, the twisted matrix $N'$ gives the geometric phase which is identical with the findings for the second choice $k=1$.The geometric phase visualizes in both cases the circular birefringence of the medium.

 Without considering the double rotation of light vector from $\varepsilon\longrightarrow\varepsilon'\longrightarrow\varepsilon"$,if we consider only the interaction of light ray $\varepsilon'$ in eq.(26) with the optical medium $N'$ in eq.(27) the outcome of the calculation is
\begin{eqnarray}
&&{{\varepsilon'}_1}^*{n'}_2{\varepsilon'}_2+{\varepsilon_2}^* {n'}_3 {\varepsilon'}_1\nonumber\\
&=&(-i e^{-i\phi}\cos\theta+\sin\theta)(k-e^{i\phi})(-ie^{i\phi}\sin\theta+\cos\theta)\nonumber \\
&+&(ie^{-i\phi}\sin\theta+\cos\theta)(e^{-i\phi}-k)(ie^{i\phi}\cos\theta+\sin\theta)\nonumber\\
&=& i[2-2ik\cos\phi-(1-\cos2\theta)(1-\cos2\phi)]
\end{eqnarray}
As a result one can finally find the phase $\Upsilon'$ for the initial adjustment of the optical medium at $k=1$
\begin{equation}
\Upsilon'= -i[2-2ik\cos\phi-(1-\cos2\theta)(1-\cos2\phi)]
\end{equation}
Thus for the specific twist of the medium for $k=1$, the geometric phase will be obtained from the difference of (39) and (36).
\begin{equation}
\Upsilon'-\Upsilon = 2i(k\cos\phi) + (1-\cos2\theta)(1-\cos 2\phi)]
\end{equation}
Hence it is seen that the GP appeared here eq.(40) is different from eq (37) though the twisted optical medium is same. The very cause of this difference is to consider the proper twist of the light ray passing though the medium.In all cases,the external twist of the system visualize the circular birefringence of the medium in terms of $k\cos\phi$.
It is seen from the above that the two geometric phases are in the form of the usual solid angle visualizing the circular birefringence
of the medium. If the optical medium is twisted arbitrarily, $k=d\theta/dz \neq constant$, we realize that the nature of the phase will
depend not only on the above solid angle but also on the nature of k. If $k=$periodic, a precessional type motion may be realized that could be studied further extensively.

\section{Angular momentum interpretation of polarization matrix and geometric phase}

The locus of the electric vector of polarized light when traces a circle, the light is known as circularly polarized light. When light is circularly polarized the two senses of rotation identifies two spin angular momentum (SAM) of photon given by $\sigma \hbar=\pm 1$ corresponding to left and right-handed circular polarization respectively.The photons having two spin angular momentum (SAM), can be visualized in another way \cite{paper11} by the two opposite directions of helicities.The conventional Poincare sphere is a SAM space where three kinds of polarized states are defined and where polarization changes from point to point.

Apart from the SAM, photons can also carry orbital angular momentum (OAM) arising from the inclination of the phase fronts with respect to beam's propagation axis.An important advancement was to realize the connection between topological charge and the orbital angular momentum of single photon by Allen et.al.\cite{paper19} for Laguerre-Gaussian (LG) beams with azimuthal phase $\exp(il\phi)$ for OAM $l$ per photon.For every value of $l$, there will be values of m from $-l$ to $+l$ and there shall be an infinite number of eigen states of $l$. Recently the orbital Poincare sphere has been sketched by Galves et.al. \cite{paper14}.All these helps to realize one that there will be two distinct representation of Poincare sphere by OAM and SAM respectively.

In our previous work \cite{paper11}, it has been suggested that with the change of polarization of light by some optical element,the parameter in connection with helicity changes.The behavior of chiral photon with a fixed helicity $(\pm1)$ in the polarized light is similar to a massless fermion having helicity $(\pm 1/2)$. These eigenvalues $(\pm 1/2)$ reflect the helicity of polarized photon. According to Berry \cite{paper3} photons have no magnetic moment, so it cannot be turned with a magnetic field over a closed path, but have the property of helicity to use.The above discussions help us \cite{paper10}\cite{paper11} to specify three variables $\theta,\phi$ and $\chi$ to parameterize a polarized photon. We would like to mention further that comprising these three parameters an extended Poincare sphere representation could be given considering the fixed spin vector attached at each point. In spherical geometry these $\phi$ and $\chi$ identify the operators of OAM and SAM by $i\hbar\frac{\partial}{\partial\phi}$ and $i\hbar\frac{\partial}{\partial\chi}$ respectively. The quantities $m$ and $\mu$ just represent the eigenvalues of the OAM and SAM operators.The Poincare sphere parameterized by $\theta$ and $\chi$ may be identified as the SAM sphere \cite{paper12} and comprising $\theta$ and $\phi$ parameters it is OAM sphere.

 In our first work \cite{paper11},with the spinorial representation of polarized photon  by spherical harmonics,the polarization matrix becomes
$$M\simeq \pmatrix{{Y_1}^0 & {Y_1}^1 \cr {Y_1}^-1 &{Y_1}^0\cr}$$
represented by each point on the Poincare sphere of OAM, parameterized by
the angles $\theta$ and $\phi$.
The elements of the polarization matrix ($2\times2$) are the product harmonics ${Y_1}^1$, ${Y_1}^{-1}$ and ${Y_1}^0$ for orbital angular momentum $l=1$.
 This indicates that for higher OAM states $l=2,3..$ the respective product harmonics ${Y_l}^m$ could be used.On the other hand, if we define our polarization matrices $M,N$ on the Poincare sphere parameterized by $\theta$ and$\chi$ \cite{paper12} it is a point on the SAM sphere.

In view of this here we may pointed out that the birefringent plate represented by the matrix $N=k \pmatrix{0 & -e^{i\phi} \cr e^{-i\phi} & 0 \cr}$ lye on the OAM sphere for $l=1$. Both circular and linear birefringence,could be measured by the angular momentum of polarized photon in terms of angle $\phi$.The inherent property of the medium gives a natural twist to the incident polarized light by an angle $\Omega$ as seen in eq. (12).
The uniform twist (small) of the optical medium per unit thickness is $k$, whose value has chosen  $k=0,1$ to represent the external twist of the birefringent optical medium. Intuitively $k$ behaves as $m$ because it is the eigenvalues of $N$. Hence we may consider it as the topological charge of the birefringent medium.

Geometric phase would be developed from the transfer of angular momentum in the course of variation of polarization or direction of propagation over a closed path.
The type of geometric phase acquired by the polarized photon depends only on the sphere/space where the path is traced out. The GP of Pancharatnam (PP)
 developed by cyclic change of polarizations will be helicity dependent visualized by the parameter $\chi$  \cite{paper10} \cite{paper12} on the SAM sphere.In the OAM sphere the GP is arising for mode transformations \cite{paper6}.
Here we have studied the GP in the OAM sphere with use of birefringent medium.
 Sanatamato \cite{paper20} pointed out that a birefringent plates affects the SAM and as well as OAM through some topological charge.He mentioned that a QP is an ordinary birefringent plate rotated at an angle $\alpha$ about the beam z-axis, with $\alpha$ given by $\alpha = \alpha(x, y)= arctan(y/x) = \phi$ where $\phi$ is the azimuthal angle in the x, y-plane.If the topological charge of the birefringent plate is q \cite{paper13}, the OAM of a light beam passing through such a "q-plate" (QP) changes by $\pm2q$ per photon.

 In view of this we may comment lastly that the birefringent plate represented by the matrix $N$ here, may be identified as $k-$ plate having topological charge $k$. Independent of the choice of the initial medium,we find here that a GP depends on the final topological charge $k$ of the twisted medium. The novelty of this work also lies in the appearance of the circular birefringence $k\cos\phi$ of the optical medium through the geometric phase of the polarized light.

{\bf Acknowledgement} This work has been fully supported by the Abdus Salam International Centre for Theoretical Physics (ICTP), Trietse, Italy. Correspondence with Prof.Santamato,Napoli,Italy is gratefully acknowledged. Moreover,I am thankful for the  help from Mr.Grassberger of ICTS, ICTP.


\begin{thebibliography}{99}
\bibitem{paper1}S.Pancharatnam; "Generalized theory of interference and its application", Proc. Ind. Acd. Sci: {\bf A44}, (1956) 247-262.
\bibitem{paper2}M.V.Berry; "The adiabatic phase and Pancharatnam phase for polarized light", J.Mod.Opt:{\bf 34}, (1987) 1401-1407.
\bibitem{paper3}M.V.Berry; 1986, "Adiabatic phase shifts for neutrons and photons", in
    Fundamental Aspects of quantum theory,eds.V.Gorini and A.Frigerio,Plenum,
    NATO ASI series, vol-144, 267-278.
\bibitem{paper4}R.Y.Chiao and Y.S.Wu;"Manifestations of Berry's topological phase for photons", Phys. Rev. Lett. {\bf 56}(1986) 933-936.
\bibitem{paper5}R.Y.Chiao and T.F.Jordan; "Lorentz group Berry phases in squeezed state of light" Phys.Lett.{\bf A132}(1988)77-81.
\bibitem{paper6}S.J.van Enk;"Geometric phase, transformations of gaussian light beams and angular momentum transfer",Optics Communications {\bf 102} (1993) 59-64.
\bibitem{paper7}R.C.Jones;'A New Calculus for the Treatment of Optical Systems. V, Properties of M-matrices",J.Opt.Soc.Am.{\bf 37},486(1942).
\bibitem{paper8}R.C.Jones;'A New Calculus for the Treatment of Optical Systems. VII, Properties of N-matrices",J.Opt.Soc.Am. {\bf 38}, 671-685 (1948).
\bibitem{paper9}R.M.A.Azzam;"Propagation of partially polarized light through anisotropic media with or without depoalrization: A differential $4 \times 4$ matrix calculus", J.Opt.Soc.Am.{\bf 68},(1978) 1756-1767.
\bibitem{paper10}D.Banerjee;"Polarization matrix and geometric phase"
    Phys.Rev.-{\bf E56},1129 (1997).
\bibitem{paper11}D.Banerjee; "The spinorial representation of polarized light and Berry phase",Comm.in Theo.Physics {\bf 3} 183-198 (1994).
\bibitem{paper12}D.Banerjee;"Geometric phase from a dielectric matrix",J.Opt.Soc.Am.-{\bf-23B}, 817-822 (2006).
\bibitem{paper13}R.A.Beth; "Mechanical detection and measurement of the angular momentum of light." Phys.Rev.{\bf 50},115-125 (1936).
\bibitem{paper14}E.J.Galvez,"Geometric phase associated with mode transformations of optical beams bearing orbital angular momentum"Phys.Rev.Lett.{\bf 90}(2003) 203901-4.
\bibitem{paper15}M.J.Padgett and J.Courtial;"Poincare sphere equivalent for light beams containing orbital angular momentum", Opt.Lett.{\bf 24}(1999) 430-432.
\bibitem{paper16}l.Allen,J,Courtial and M.J.Padgett; Matrix formulation for the propagation of light beams with orbital and spin angular momenta;Phys.Rev.{\bf E60}(1999),7497-7502.
\bibitem{paper17}S.C.Tiwari,"Geometric phase in optica and angular momentum of light"; J.Mod.Opt.{\bf 51} (2004) 2297-2304.
\bibitem{paper18} S.Huard,{\it Polarization of Light}(John Wiley and Sons, 1997).
\bibitem{paper19}L.Allen,M.W.Beijersbergen.R.J.C.Spreeuw and J.P.Woerdman;"Orbital angular momentum of light and the transformation of Lagurre-Gaussian laser modes",Phys.Rev.-A{\bf 45},(1992) 8185-8189.
\bibitem{paper20} E.Santamto et.al.; "Light propagation in birefringent plates with topological charge";Phys.Rev-{\bf E69}(2004)056613-7.
%\bibitem{paper21} A.Mair, A.Vaziri,G.Weihs and A.Zeilinger,"Entanglement of the orbital angular momentum states of photons";Nature {\bf %412},(2001)313-316.
\end{thebibliography}
\end{document}